# 'Missing' Women in Economics Academia: Evidence from India

Ambrish Dongre[a], Karan Singhal[a], Upasak Das[b]

[a]Indian Institute of Management, Ahmedabad, India; [b]University of Manchester, Manchester, United Kingdom

*Authors*

Ambrish Dongre (*corresponding author*),

Wing 13 F, Faculty Block, Heritage Campus,

Indian Institute of Management,

Vastrapur, Ahmedabad

Gujarat (India): 380015

ambrishd@iima.ac.in

Karan Singhal,

Indian Institute of Management,

Vastrapur, Ahmedabad

Gujarat (India): 380015

karansinghal1993@gmail.com

Upasak Das,

Presidential Fellow,

University of Manchester

upasak.das@manchester.ac.uk

**Acknowledgements**

We would like to thank Anindya Chakrabarty, Vijaya Sherry Chand, Pritha Dev, Sahil Gandhi, Maitreesh Ghatak, Reetika Khera, Abhiroop Mukhopadhyay, Jeevant Rampal, Kavitha Ranganathan, Ankur Sarin, Chinmay Tumbe, Jeemol Unni, and Nisha Vernekar for discussions and helpful comments. We are also grateful to Vamsi Antyakula, Sai Shruthi Balaji, Nawaz NM, Ishika Saha and Soham Shevde for research assistance. Finally, we would like to thank all the individuals who gave their valuable time and responded candidly to our questions.

**Abstract**

This paper documents the representation of women in Economics academia in India by analysing the share of women in faculty positions, in journal publication and their participation in a prestigious conference held annually since 2004. Data from 120 elite institutions shows that only 28.5 percent of the Economics faculty members are women. Of the authors of more than 1300 papers mentioned in the final schedule of a prestigious annual research conference, women constitute only 29 percent of authors over the period 2004 to 2017. Of the authors whose research was published in a leading India-focussed Economic journal over the same period, 26 percent were women. These figures are indeed low since women are not under-represented at the master's level. Further, the proportion of women in doctoral programmes has increased over time, and is now almost 50 percent. Tendency of women who earn a doctorate abroad, to not return to India, and responsibilities towards the family post marriage or child-birth may partly explain such under-representation of women in Economics academia.

## 1. Introduction

India, like rest of the developing world, has witnessed tremendous progress in women's education. Gender gaps in enrolments at elementary levels have disappeared while gaps at the secondary and higher education levels have narrowed considerably (ASER 2018; NSS 2019). But these increased educational achievements haven't translated into corresponding labour market outcomes. Contrary to the rest of the world, participation of women in the labour force has gone down in the last three decades, and is now one of the lowest among emerging markets and developing countries (Klasen and Pieters 2015; Afridi et al. 2018). The gender gap is also evident in wages, presence in prestigious and highly paid professions, entrepreneurship, and leadership positions. This has pushed India to be one of the lowest ranked countries as per the latest Global Gender Gap Report (WEF 2020).

This paper documents the presence of women in Economics academia in India, a niche labour market. This is important for several reasons. Not much is known about presence of women and other disadvantaged groups in Economics or in other disciplines in India. The official documents merely report aggregate faculty numbers, and don't provide discipline-wise and social group-wise faculty breakdown. There are only a few studies that explore career trajectories of women in academia, and that too are focused on Engineering, Science and Medicine (Chanana 2003; Gupta 2007, 2016). But why should we care about presence of women and other disadvantaged groups in academia? It is being gradually accepted that the background of the researchers matter in what is studied and how it is studied. As a result, a discipline with people from various social backgrounds is likely to ask different questions, study those questions from different perspectives, and arrive at different conclusions. For example, May et al. (2014) has documented differences in the views of male and female members of the American Economic Association (AEA) on several issues such as minimum wages, labour standards, health insurance, explanations of the gender wage gap, and issues around equal opportunity in the labour market. Thus, presence of individuals with diverse backgrounds makes a discipline more robust, relevant and dynamic. This is especially true for Social Sciences such as Economics which aim to understand human behaviour and impact public policy (Mester 2019).

There is an international context to this research as well. There has been intense debate within Economics academia about stagnant share of women faculty over time relative to other social Sciences (Ginther et al. 2017; Aurio et al. 2019; Lundberg and Stearns 2019). The debate has highlighted various institutional and non-institutional factors causing women to drop at various stages of an academic ladder, the phenomenon referred to as 'leaky pipeline' (Bayer and Rouse 2016; Wu 2018; AEA 2019). But the debate and evidence generated has almost exclusively focused on the United States and Europe (Lundberg 2020). The contexts in which academia operates in developing countries are different from that of the developed world and hence, extent of presence of women, reasons for their differential experiences, and therefore potential solutions are also likely to be very different.

Drawing from the literature, this paper focuses on three dimensions of being in Economics academia in India. First dimension is, being a regular, full-time faculty in an 'elite' institution offering a postgraduate degree or diploma. Data required to do so has been meticulously collected from the official websites of 120 'elite' institutions between May 2019 and February 2020. Reasons for focusing on 'elite' institutions are explained later. The second dimension that we measure is attending and presenting in one of the most prestigious research conferences in India, namely the *Annual Conference on Growth and Development*, hosted by the Indian Statistical Institute, Delhi (ISI-D), annually since 2004. Conferences are critical for visibility, professional exchange, networking and collaborations, and therefore an important aspect of being an academic (Eden 2016; Leon and McQuillin 2018; Henderson and Burford 2019; Lipton 2019). We have manually compiled individual author-level data of more than 1300 papers presented at the ISI conference since its beginning in 2004, until 2017. The lack of access to similar data on other conferences despite our efforts prevented us from including more conferences in our analysis. The third dimension is publication of research articles in journals. For the reasons explained later, we have chosen to focus on articles published in the Indian Journal of Labour Economics (IJLE), which predominantly publishes academic work of economists based in India.

What are the main results? Analysing data from elite institutions that employ at least one faculty member with a PhD in Economics, we find that overall, 28.5 per cent faculty members are women. The share is 22.7 per cent at Professor level, 32.5 per cent at Associate level, and 32.2 per cent at Assistant Professor level. These overall figures hide substantial heterogeneity across institutions. For example, private institutions and institutions that are primarily financed by sub-national governments (referred to as state governments in the Indian context) have relatively higher share of women faculty compared to other types of institutions (details in the next section). Data from the ISI conference reveals that women constitute only 29 per cent of the authors of papers presented at the conference for the period 2004 to 2017, and shows no improvement over the time-period under consideration. When it comes to journal publications, only 26.4 percent of the authors whose work was published in IJLE during 2004 to 2017 were woman. This fraction has improved, and moved marginally above 30 per cent during 2011 to 2017 as compared 24 per cent during 2004 to 2010.

Is this share low? We think it is. Using data obtained from various official sources, we show that women are *not* under-represented at the master's level. In fact, they constitute more than half of the students at the master's level. The share of women drops at the PhD level as compared to the master's, but it has been increasing steadily over time and is now approaching 50 per cent. This stands in contrast to the United States, where the percentage of women at the master's and the doctoral levels has been less than 35 per cent (Buckles 2019), and makes the Indian case quite unique. Thus, the low share of women among the Economics academia in India can be better understood if we can unpack the following two stages: a) transition from a master's degree to a PhD, and more importantly, b) transition from being a PhD holder to having a faculty position or conducting research.

We attempt to do so through two ways. First, based on the information available in the public domain, we analyse career choices of the alumni of one of the most prestigious Economics master's programmes in India. It reveals an interesting fact: women are more likely to continue staying abroad, conditional on obtaining the doctoral degree outside India. A large number of faculties in elite institutions have doctorates from institutions in the United States and, to some extent, Europe. Therefore, women's lower probability to return to India post their doctoral studies could be an important variable in keeping the number of women faculty low at least in elite institutions. To our knowledge, this factor has not been highlighted in the literature, and is probably unique to the developing countries. Secondly, we conducted detailed interviews with a selective and small sample of women who have obtained a postgraduate degree in Economics from prestigious institutions in India. These conversations suggest that the time taken to complete the doctoral education (typically four to six years) and the implication it may have on age at marriage is an important consideration, especially in Indian society where there are strong notions about the appropriate marriage age for women. Responsibilities towards the family, especially that of raising children, make it difficult for women to pursue rigorous research either on their own or through collaborations, which then have negative implications on conference participation. Thus, gender norms do play a role in reducing women's labour force participation, something which has been highlighted elsewhere as well.

Other explanations for the lower share of women in Economics academia in India can't be ruled out. Do women prefer a non-academic job (say private corporate sector or non-governmental organisations) over an academic job post their master's or PhD? Is it socialisation and enculturation during the master's or doctoral programme which is at play as has been found in the context of STEM institutions in India (Gupta 2007)? Does the quality of doctoral research vary across gender? Do men and women research different topics, which might have implications for job-market opportunities? Are there biases in recruitment processes? Is the 'work culture' to blame? Is the conference participation low due to conferences not being 'family friendly' (Bos et al. 2017)? Exploring the role of these factors in the context of Economics academia will help us designing appropriate policies and processes to improve presence of women, and hence remains an important research agenda.

The rest of this paper proceeds as follows. The next section describes the data collected and analysed. Main findings are in section 3, followed by a discussion in section 4. Section 5 concludes.

2. **Data**

*Women among faculty*

A major challenge in determining share of women among Economics faculty is that such data doesn't exist, at least in the public domain. Thus, to keep our enterprise feasible and time-bound, we couldn't include all higher education institutions in our analysis. We limited ourselves to 'elite' institutions that award postgraduate degree or diploma. These institutions have updated websites from which information about the courses offered and the faculty can be obtained.

For the purpose of this paper, we define 'elite' institutions as the institutions that receive a high place in an annual ranking exercise conducted by the Ministry of Human Development (MHRD), a federal ministry in charge of policymaking and financing of higher education in the country. This exercise, known as National Institution Ranking Framework (NIRF) ranks institutions in nine categories: *Overall*, *Universities*, *Engineering*, *Colleges*, *Management*, *Pharmacy*, *Law*, *Architecture* and *Medical*. *Overall* was introduced as a category in 2019, where any institution, independent of its discipline, was given an *overall* rank if the institution had at least 100 students at the undergraduate and master's levels. The broad parameters on which these rankings are based, are common across the nine categories, and are as follows:

a) Teaching, learning and resources
b) Research and professional practice
c) Graduation outcomes
d) Outreach and inclusivity
e) Perception

These parameters are given specific weights. Each parameter consists of sub-heads with specific weights too. Table A1 in the online annexure illustrates this for *Overall* category. We focus on institutions in three categories — *Overall*, *Universities* and *Management*, and include in our analysis, institutions that are ranked in the top 50 in *Management* category, and institutions that are ranked in the top 100 in *Universities* and *Overall* categories.

After identifying the institutions, the next step was to check whether the institution had any regular full-time faculty member with a PhD in Economics. By excluding the institutions that don't have such a faculty yielded a sample of 56 institutions that are listed in the top 100 in the *Overall* category, 52 institutions that are listed in the top 100 in the *Universities* category, and 33 institutions that are listed in the top 50 in the *Management* category. The institutions that are excluded are typically institutions devoted to medical sciences, physical and natural sciences, and engineering.

The NIRF rankings are not specific to Economics. To complement it, we also refer to the rankings of institutions by *Research Papers in Economics*, popularly known as RePEc[i]. Out of the 225 Indian institutions that are listed on RePEc, we focus on the top 25 per cent as of January 2020[ii]. These rankings are based on the number of (distinct) research works weighted by impact factors[iii]. As a result, the list also includes institutions which are solely focused on research and don't offer any degree at the postgraduate level. Excluding such institutions yields a final sample of 39 institutions. Combining NIRF and RePEc rankings yields a sample of 120 institutions, i.e. the institutions that are ranked in at least one of the four lists — NIRF top 50 in *Management,* NIRF top 100 in *Universities* and NIRF top 100 in *Overall*, and RePEc top 25 per cent[iv].

Once an institution has been identified, we identify individuals with doctoral degrees in Economics, among the regular full-time faculty members. We exclude visiting or adjunct faculty as well as honorary faculty. An individual with a doctorate in Economics can belong to a department other than Economics too, thus extending the search beyond Economics department[v]. Further, in many instances, details of the faculty's educational background on the institution website doesn't explicitly mention whether they have a PhD in Economics. Therefore, we had to look up their personal website, professional profile or CV as well. We have not included those individuals whose specialisation at the doctorate level was unclear.

Though the sample might appear restrictive due to it being limited to elite institutions, the institutions in the sample are spread across multiple states in India (Table 1). Further, these institutions span different regulatory regimes (Table 2). They differ in terms of who established them (an act by the federal or sub-national (state) government), their funding (fully or partly by government or self-funded), course offerings (full-fledged university with several departments or an institution offering degree or diploma in a specific discipline), recruitment policies, and degree granting powers, just to mention a few examples[vi].

*Women in ISI Conference*

As mentioned before, the *Annual Conference in Growth and Development* (held at ISI-Delhi) is a prestigious conference attended by individuals affiliated with well-known institutions in India and abroad, and includes students, faculty and researchers.

The procedure of manuscript selection at this conference is similar to that of other conferences. The call for papers is announced, the authors submit their manuscripts, the conference organisers inform the authors the status of their submissions, and finally, the authors of the 'accepted' manuscripts confirm their attendance and submit an updated version of the manuscript. Some of the waitlisted submissions might move to the selected list if the authors of the originally selected manuscripts convey their inability to attend the conference. Finally, the conference schedule is uploaded on the conference webpage, and includes details such as paper title, author name(s), and institutional affiliation(s) [vii]. We downloaded these details for all the years in which the conference was held, i.e. 2004 to 2017. Thus, what we have is the final schedule. We complemented the available details with information on the authors' gender and whether their institution of affiliation was located in India. These were obtained through profiles and CVs available on personal websites, institutional websites, and in some cases, LinkedIn profiles. We would have preferred to have a list of submitted manuscripts, and information on who dropped out before the final schedule was frozen, neither of which was available[viii].

*Women in journal publication*

We initially focussed on two journals – The Indian Journal of Labour Economics (IJLE), and Indian Journal of Agricultural Economics (IJAE). These journals predominantly publish work by the economists based in India. While Economic & Political Weekly (EPW) is also another journal that meets similar criteria, we did not include it because it is a multi-disciplinary journal, and as a result, large fraction of authors specialises in subjects other than Economics. Further, not all articles in EPW are peer-reviewed.

We obtained the list of articles published in IJLE and IJAE during 2004 to 2017 (both, inclusive) from the SCOPUS database. The next step, determining the gender of the authors, proved time-consuming and difficult, especially in case of IJAE. Even after extensive online search that included institutions that the authors were affiliated as per the journal records, and various other websites (such as Google Scholar, LinkedIn etc.), we were unable to determine gender of 36.5 per cent of the IJAE authors. Given the significant extent of missing information, we have decided not to report findings from IJAE.

*Women in postgraduate programmes*

Aggregate information on the number and share of women in Master's, Master of Philosophy (MPhil), and Doctor of Philosophy (PhD or DPhil) programmes has been obtained from various reports of the *All India Survey of Higher Education* (AISHE), an official report put out by the Department of Higher Education of the MHRD[ix]. Information on the share of women among students in selected prestigious institutions has been obtained through queries under the Right to Information (RTI) Act, links available on the institute websites, and placement brochures which are available in the public domain.

*Tracking alumni*

Master's in Quantitative Economics (MSQE) is one of the most selective and prestigious master's program in Economics in India. It is offered by the Indian Statistical Institute (ISI) through its Delhi and Kolkata centres. 336 individuals graduated from this programme between 1998 and 2013 i.e. roughly 21 individuals per year. We obtained the names of these individuals through the link on ISI's website[x]. We then searched for these names online and obtained information on their current location, whether they have earned a PhD or are currently enrolled in a doctoral programme, and the location of their PhD. We were able to find this information for 287 individuals (85.41 per cent).

3. Findings

*Women among faculty members*

The percentage of women among faculty members with a doctorate in Economics across all institutions that we have covered is 29.6 per cent (Table 3). It varies between 28 per cent and 32 per cent across the four lists of elite institutions that are in our sample. The share of women faculty that we find in these Indian institutions is higher than what is found in the US, and a number of European countries (Lundberg 2018; Auriol et al. 2019). In all the lists, the percentage is lowest at the Professor level and higher at Assistant and Associate levels. This is consistent with what has been found in other contexts. Interestingly, there is not much difference between the percentage of women at the Assistant Professor and the Associate Professor levels.

Does the share of women faculty vary across types of institutions? Table 4 shows that Central Universities, Institutions of National Importance, Institutions recognised by the universities to offer a postgraduate degree, and Indian Institute(s) of Management (IIMs) have a less than 25 per cent share of women among their faculty members, while private universities have close to 50 per cent. Women faculty constitute roughly one-third of the total faculty strength of stand-alone institutions other than IIMs, 'Deemed to be Universities' and State Universities. We are not aware of any literature that has either documented differences in the share of women across types of institutions in India and more importantly, reasons for such differences. Understanding the causes of these differences should be an important research and policy agenda.

*Women in the ISI Conference*

The share of women authors among the authors of accepted papers is around 30 per cent after 2006 and has not changed over time (Figure 1). Another way to look at it is Figure 2 which shows that there are over 1.4 men per paper against an average of 0.6 women per paper[xi].

Figure 3 presents the composition of authors disaggregated into three categories: papers with (i) all women authors (one or multiple), (ii) all men authors (one or multiple), and (iii) a mix of men and women authors (at least one male and one female author). We find that a major share of the papers presented is authored by only men and this remains unchanged over time. The share of papers with all women authors is less than 20 per cent and does not show signs of an increase over time. The share of papers with both men and women authors has remained unchanged at around 30 per cent. Findings are similar if we consider only co-authored papers (i.e. at least two authors)[xii].

*Women in IJLE*

The share of women authors is 26.4 per cent over the entire period 2004 to 2017. The share has increased from 24.4 per cent between 2004 to 2010 to 31 per cent between 2011 to 2017. Figure 4 presents composition of only women, only men and 'mixed' authors for IJLE papers similar to composition of ISI conference paper authors shown in Figure 3. We find that a major share of published papers are authored by only men but this has substantially varied over time ranging from 48.5 percent (in 2018) to 83.3 percent (in 2011), and shows signs of a decrease in the most recent years (56.8 percent for 2015-19 compared to 69.8 percent for 2004-08). Share of papers with mixed authors too has substantially varied over time (3.8 percent to 29.4 percent) and has seen an increase in most recent years, more than doubling from 10.6 percent in 2004-08 to 21.6 percent in 2015-19. Share of papers with only female authors has also varied like others (10 to 33 percent) but does not show any consistent sign of an increase/improvement.

4. **Discussion**

In this section, we explore potential explanations for the share of women being less than one-third among the faculty members as well as among the presenters at the ISI conference.

*Women in Master's Programmes in Economics*

One reason for lower presence of women in academia could be their lower presence in master's programme. Does that apply in the Indian context? Table 5 depicts the percentage of women who were enrolled in and completed a master's programme in Economics, across India. This number has been above 50 per cent since the beginning of this decade, and has continued to increase. Thus, women have been outnumbering men in postgraduate Economics programmes overall.

Is this trend similar in some of the most prestigious institutions offering a master's in Economics where entry is quite competitive, the course tends to be more mathematical in nature, and rigorous? We have collected the data for ISI — Delhi and Kolkata (combined), Delhi School of Economics (DSE), Centre for Development Studies (CDS), Indira Gandhi Institute of Development Research (IGIDR) and Centre for International Trade and Development at Jawaharlal Nehru University (CITD-JNU) [xiii]. Numbers reported in Tables 6 to 9 reveal that though there are fluctuations from year to year, the percentage of women is close to or even above 50 per cent with the exception of ISI, where the share for the entire period is 41 per cent. These figures are in contrast to evidence from the US and some European countries where the representation of women in Economics is lower at the postgraduate and doctoral levels.

*Women in MPhil. and PhD programmes in Economics*

Next, we examine the share of women in MPhil and PhD programmes. Data from the MHRD suggests that the percentage of women enrolling and earning an MPhil has been almost 50 per cent for a decade, and it has continued to increase even further, reaching marginally above 60 per cent (Table 10). But it is the PhD programme where a dramatic decline in the percentage of women can be seen. Till 2014-15, around 40 per cent of those who enrolled in and (and also earned) the PhD were women. Recall that the percentage of women at the master's level was above 50 per cent during this time. Thus, roughly, there is a 10-percentage point gap between the proportion of women at the master's and the PhD levels. Even though the share of women has improved and has reached closer to 50 per cent at the PhD level, it is lower than the fraction of women enrolled in the master's programmes[xiv].

A number of individuals go abroad to pursue a doctorate in Economics. Table 11 shows the percentage of women among Indian citizens who earned doctorates in Economics in the US, probably the most important destination outside India for those seeking a doctoral degree in Economics. The percentage has always been below 50 per cent till 2010. In fact, the share of women among Indian citizens earning a doctorate in Economics for the period 1997-2010 as a whole is 40.3 per cent. For the period 2011-2017, the share has increased to 50.4 per cent.

Thus, the share of women at the master's and the PhD levels is much higher than the overall share of women among the faculty members, and importantly, higher than the share of women at the Assistant Professor level, the next step on academic ladder for a PhD holder.

*Share of female faculty across locations within India*

Does the location of an institution play a role in impacting the share of women in that institution? One plausible hypothesis could be that women who earn doctoral degree in Economics would prefer to be located in cities with more job opportunities. Alternatively, they might prefer to marry someone with appropriate levels of education, and certain types of occupation. If this is the case, then we should expect share of women faculty much higher in bigger cities than the smaller ones.

We compare the share of women in institutions which are located in six of India's largest cities, with the rest[xv]. Results in Table 12 show that on average, the share of women among faculty members is higher in metros than other cities. What is interesting is that despite the rising share at the PhD level, the share of women has actually gone down (in metros) or has increased only marginally (non-metro locations) at Assistant Professor level compared to Associate Professor level. Further, Table A3 in the online annexure shows substantial variation even within the metros. Chennai and Hyderabad have a much lower share of women faculty compared to not just other metros but also non-metro locations. Thus, share of women among faculty is low irrespective of locations.

*Preference to settle abroad versus in India*

As mentioned earlier, we were able to obtain information for 287 out of 336 alumni of the ISI MSQE programme who graduated between 1998 and 2013. Of these 287, 41 per cent are women, and 38.7 per cent have earned a PhD. Interestingly, there is no difference in the fraction of PhDs between male and female alumni. Of those who have already earned their PhD, only 10.8 per cent have earned it in India. Thus, MSQE graduates overwhelmingly pursue PhD abroad.

What is important for our purpose is that, of those who completed their PhD abroad, 42.9 per cent of male alumni and only 24.4 per cent of female alumni are currently in India. The findings could be extrapolated to other selective institutions as well (such as Delhi School of Economics), and therefore, suggests another reason why the share of women faculty could be low, especially in elite institutions. To our knowledge, this factor is likely to be relevant to other developing countries as well, and is not highlighted in literature.

*Findings from semi-structured interviews with women*

To understand what factors are at play in reducing the percentage of women in at the PhD level and then in academia, as compared to their presence at the master's level, we interviewed 12 women in the age-group 25 to 35 years, and who have graduated from prestigious institutions in the country. These were (i) women with a PhD in Economics from an elite Indian institute and currently *working* as faculty in India, (ii) women who *did not opt for a PhD*, and (iii) women with a PhD and currently *not working* as faculty. We summarise the findings below.

One of our respondents from the first category mentioned that she did not face any problems from her own parents or in-laws while pursuing her PhD. However, she found it difficult to pursue research work rigorously because of her responsibilities as a mother.[xvi] This, in some way, hindered collaboration with other researchers as well. More often than not, she was unable to fully commit in collaborative efforts that might require online meetings during non-office hours, which in turn, affected the quality time spent with her child. She had missed conferences despite securing funds to travel for the same reason. Notably, she was among the top performers in her class at the postgraduate level.

Our analysis from the interviews for the second category revealed that the time required to complete a doctoral programme can discourage some women. Even though some of the respondents wanted to pursue a PhD, they found it difficult to convince family and relatives about the opportunities that may open up post-PhD. Further, there was worry about their marriages getting delayed, which added to their families' dissuasion. Another respondent explained that she was keen to take up admission outside India, specifically in institutions in the US or Europe. She felt that these countries were safer and more "equal," had better opportunities for women, and also that the move would help her "move far away from the pressures of marriage", echoing the general concerns aired by other respondents too.

One of the respondents who opted not to work after the completion of her PhD (third category) said her husband wanted her to take care of their children. Since they were economically better off, she would not need to work regularly. However, she was free to take periodic part-time/short-term teaching assignments, while still devoting much of her time to her kids.

## 5. Conclusion

To summarise, there is limited evidence on the presence of women in academia in developing countries including India. We fill this gap through systematic and intensive data collection on the share of women among the Economics faculty in 120 elite institutions that are spread across the country and span different regulatory regimes in the country. We complement this by documenting the share of women participating in a prestigious research conference.

Data shows that only 28.5 per cent of the Economics faculty members are women. The percentage is lowest at the Professor level and higher at Assistant and Associate Professor levels. The share of women faculty varies across institutions and locations. Though the share might look higher when compared to the US and some European countries, it is low given that women constitute at least 50 per cent of students at the master's level. The first drop in women's share occurs at the PhD level. Data suggests that women's share in PhD is improving over time and now inching towards the halfway mark. But that is not yet reflected in faculty share. This is similar to what is observed in Science disciplines and scientific research institutes in India (Gupta 2016). So the real issue in the Indian context is how to retain women in Economics academia. Compilation and analysis of data of authors of more than 1300 papers which were presented at the prestigious research conference reveals that women constitute only 29 per cent of authors in the final schedule. Tracking the alumni of a highly selective master's programme and conversations with a non-random sample of women suggest that the time taken to complete a doctoral programme, responsibilities towards family post-marriage, preference to stay abroad post-PhD conditional on earning PhD outside India play a role in keeping the share of women faculty low.

How does one proceed from here? It would have to be multi-pronged effort with contributions from several actors – government, institutions, and not to mention, individual academics. The first step should be collection, and easy access to discipline-wise data on presence of women and other disadvantaged groups. As of today, we don't even know if the presence of women in Economics academia is lower or higher than say a decade ago. The government already collects data from all the institutions in the country. But its annual reports provide only aggregate figures. It would be more useful if it can be disaggregated and presented at the level of discipline. Alternatively, the government should make the data freely available on its website so that researchers can access it. Analysis of such data will inform us about where we stand and the direction in which we are we moving. The same applies to the conference organisers as well. We could analyse the ISI Conference data because the conference schedules right from the first year are available online. That wasn't the case with other important conferences that are organised in India.

Making conferences more inclusive might be a bit easier, keeping in mind the constraints that conference organizers face[xvii]. Providing information about the child-care facilities at or near the venue, designing schedules in such a way that time between two obligations is minimised, virtual conferencing options at least for some participants are some of the ideas which have worked in one context or the other (Bos et al. 2017)[xviii].

Role of various institutional or organisational factors is quite important. Evidence from STEM disciplines in India suggest that hierarchical culture within the organisation, lack of time for women to socialise beyond office hours, their absence from informal networks, they not being 'visible' and 'well-connected', lack of support from the superiors, 'hidden' social exclusion, dual burden (office and home responsibilities) as well as the tendency of authorities to use 'dual burden' as an excuse to not recruit, promote or give additional responsibilities to women make navigating the career path more difficult (Chanana 2003; Gupta 2007, 2016; Kurup et al. 2010; Sabharwal et al. 2019). Since Economics departments are part of the same sociocultural milieu, these factors are likely to matter to the career trajectory and experiences of women in Economics academia as well. So any step that eliminates or mitigates the impact of these factors would improve organisational environment, and prove beneficial to all, and not just to women. In addition, wide circulation of job announcements, actively reaching out to qualified women individuals, fair recruitment practices and performance reviews, exploring possibility of employment of spouses in the same organization or institutions within the same city keeping regulatory aspects in mind, campus housing, crèche and elderly care facilities on the campus, flexible work timings, research and travel grants, dissemination of important information through formal means and not through informal channels will go a long way in making the workplace more inclusive.

Economics academia also need to undertake efforts to make the discipline more inclusive. Example of the American Economic Association (AEA) is quite instructive. The AEA has, on its website, a section which provides concrete suggestions on how to build "…*a more diverse, inclusive and productive profession*"[xix]. The AEA also has a *Committee on the Status of Women in the Economic Profession* which is charged with monitoring the progress of women economists and undertaking activities that help women promote their careers. It conducts surveys, organises professional development and mentoring workshops, facilitate internship programs and various other activities for women in Economics, including undergraduate students. The Indian Economic Association (or other organisations representing the profession) can initiate the activities suitable to the Indian context.

Academics themselves can play a powerful role in shaping careers of their students in academia. They can inform students about the doctoral studies and prospects thereafter. They can institute mechanisms to connect current students to the seniors either in doctoral programmes or at the level of assistant professors, with whom students might be able to relate better. The period of doctoral work which lasts three to four years is a crucial period. In addition to the quality of doctoral work, academics influence how individuals transform from being students to researchers as they learn to deal with indeterminate and open-ended nature of any research. By facilitating healthy work environment, nudging doctoral students to develop written and oral communication skills, encouraging them to attend and present in conferences, directing them to relevant online and offline resources, academics can help doctoral students especially those who are more disadvantaged, to develop skills and confidence to navigate their careers.

# References

AEA Professional Climate Survey: Final Report, 2019. American Economic Association.

Afridi, F., Dinkelman, T., Mahajan, K., 2018. Why are fewer married women joining the work force in rural India? A decomposition analysis over two decades. *Journal of Population Economics* 31, 783–818. https://doi.org/10.1007/s00148-017-0671-y

All India Survey on Higher Education (AISHE) 2018-19, 2019. Department of Higher Education, Ministry of Human Resource Development, Government of India. Available at http://aishe.nic.in/aishe/reports

Annual Status of Education Report (ASER) 2018 (Rural), 2019. ASER Centre, New Delhi. http://www.asercentre.org/Keywords/p/337.html

Auriol, E., Friebel, G. and Wilhelm, S., 2019. Women in European Economics. *Mimeo*.

Bayer, A., Rouse, C.E., 2016. Diversity in the Economics Profession: A New Attack on an Old Problem. *Journal of Economic Perspectives* 30, 221–42. https://doi.org/10.1257/jep.30.4.221

Bos, A.L., Sweet-Cushman, J. and Schneider, M.C., 2019. Family-friendly academic conferences: a missing link to fix the "leaky pipeline"? *Politics, Groups, and Identities*, 7(3), pp.748-758.

Buckles, K., 2019. Fixing the Leaky Pipeline: Strategies for Making Economics Work for

Women at Every Stage. *Journal of Economic Perspectives* 33, 43–60. https://doi.org/10.1257/jep.33.1.43

Chanana, K., 2003. Visibility, Gender and the Careers of Women Faculty in an Indian University. *McGill Journal of Education*, 38(3). https://mje.mcgill.ca/article/view/8701

Eden, D., 2016. Women's participation in academic conferences in Israel, *Journal of Higher Education Policy and Management*, 38:4, 406-421, DOI:10.1080/1360080X.2016.1181887

Ginther, D.K., Kahn, S., McCloskey, J., 2017. Gender and Academics, in: The New Palgrave Dictionary of Economics. Palgrave Macmillan UK, London, pp. 1–18. https://doi.org/10.1057/978-1-349-95121-5_3039-1

Gupta, N., 2007. Indian Women in Doctoral Education in Science and Engineering: A Study of Informal Milieu at the Reputed Indian Institutes of Technology. *Science, Technology & Human Values*, 32(5), 507-533. https://doi.org/10.1177/0895904805303200

Gupta, N., 2016. Perceptions of the Work Environment: The Issue of Gender in Indian Scientific Research Institutes. *Indian Journal of Gender Studies*, 23(3), 437-466. https://doi.org/10.1177/0971521516656079

Henderson, E.F., Burford, J., 2020. Thoughtful gatherings: gendering conferences as spaces of learning, knowledge production and community. Gender and Education 32, 1–10. https://doi.org/10.1080/09540253.2019.1691718

Kurup, A., Maithreyi, R., Kantharaju, B., Godbole, R. (2010). Trained Scientific Women Power: How Much are We Losing and Why? *Research Report by Indian Academy of Sciences and National Institute of Advanced Studies*. http://eprints.nias.res.in/142/1/IAS-NIAS-Report.pdf

Klasen, S., Pieters, J., 2015. What Explains the Stagnation of Female Labor Force Participation in Urban India? *The World Bank Economic Review* 29, 449–478. https://doi.org/10.1093/wber/lhv003

Leon, Fernanda L. L. de, and Ben McQuillin. 2018. "The Role of Conferences on the Pathway to Academic Impact: Evidence from a Natural Experiment." *Journal of Human Resources*, September, 1116-8387R. https://doi.org/10.3368/jhr.55.1.1116-8387R.

Lipton, B., 2019. Conference baby: Gendered bodies, knowledge, and re/turning to academia. *Qualitative Inquiry*, 25(2), pp.160-162.

Lundberg, S., 2018. Report: Committee on the Status of Women in the Economics Profession (CSWEP). *AEA Papers and Proceedings*. 108: 704-721. https://doi.org/10.1257/pandp.108.704

Lundberg, S., 2020. Women in Economics. A VoxEU.org book. Available at

https://voxeu.org/article/women-economics-profession-new-ebook

Lundberg, S., Stearns, J., 2019. Women in Economics: Stalled Progress. *Journal of Economic Perspectives* 33, 3–22. https://doi.org/10.1257/jep.33.1.3

May, A.M., McGarvey, M.G., Whaples, R., 2014. Are Disagreements Among Male and Female Economists Marginal at Best?: A Survey of AEA Members and Their Views on Economics and Economic Policy. *Contemporary Economic Policy* 32, 111–132. https://doi.org/10.1111/coep.12004

Mester, L., 2019. Increasing Diversity, Inclusion, and Opportunity in Economics: Perspectives of a Brown-Eyed Economist. Presentation made at *Second Annual Women in Economics Symposium*, Federal Reserve Bank of St. Louis. https://www.stlouisfed.org/education/women-in-economics/2019/mester-presentation

National Sample Survey (NSS). 2019. Key Indicators of Household Social Consumption on Education in India. NSS 75th Round (July 2017-June 2018), National Statistical Office, Ministry of Statistics & Programme Implementation, Government of India. http://mospi.gov.in/sites/default/files/publication_reports/KI_Education_75th_Final.pdf

Sabharwal, N.S., Henderson, E.F., Joseph, R.S., 2019. Hidden social exclusion in Indian academia: gender, caste and conference participation. Gender and Education 1–16. https://doi.org/10.1080/09540253.2019.1685657

Status of Women in Science among Select Institutions in India: Policy Implications, 2017. Society for Socio-Economic Studies and Services (SSESS), Kolkata; Supported by NITI Aayog, Government of India. Available at https://niti.gov.in/writereaddata/files/document_publication/Final_Report_Women_In_Science_SSESS.pdf

World Economic Forum (WEF), 2020. Report: Global Gender Gap Report 2020. Available at http://www3.weforum.org/docs/WEF_GGGR_2020.pdf

Wu, A.H., 2018. Gendered Language on the Economics Job Market Rumors Forum. *AEA Papers and Proceedings* 108, 175–79. https://doi.org/10.1257/pandp.20181101

**Figures**

Figure 1: Share of female authors among all authors

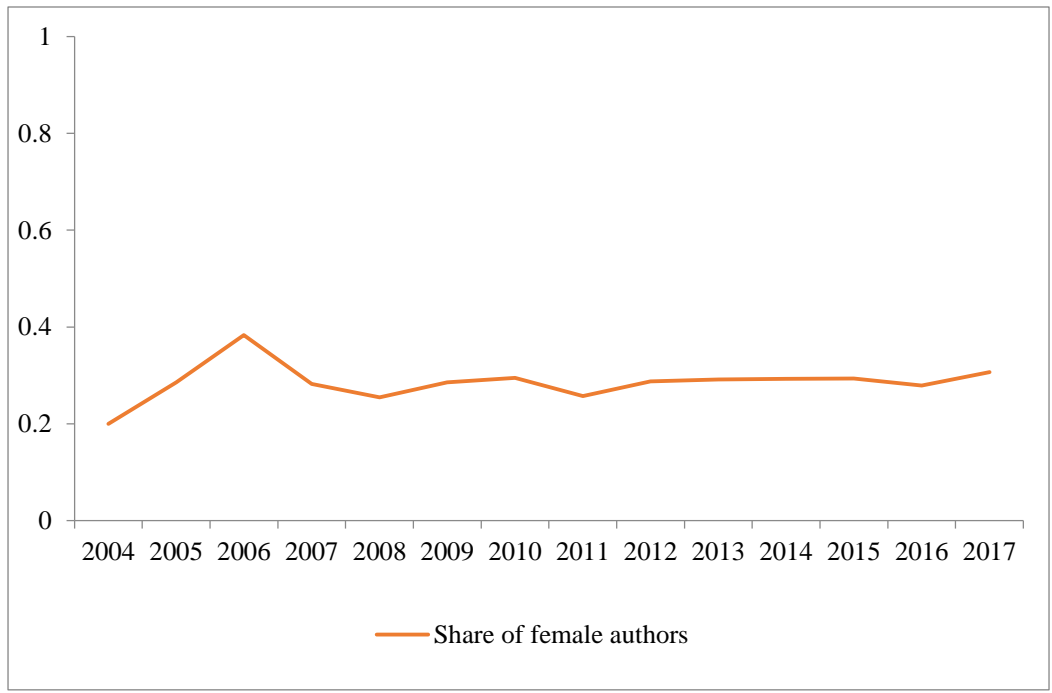

*Source:* Annual Conference on Economic Growth and Development at ISI, Delhi from 2004 to 2017

Figure 2: Total number of authors per paper across gender

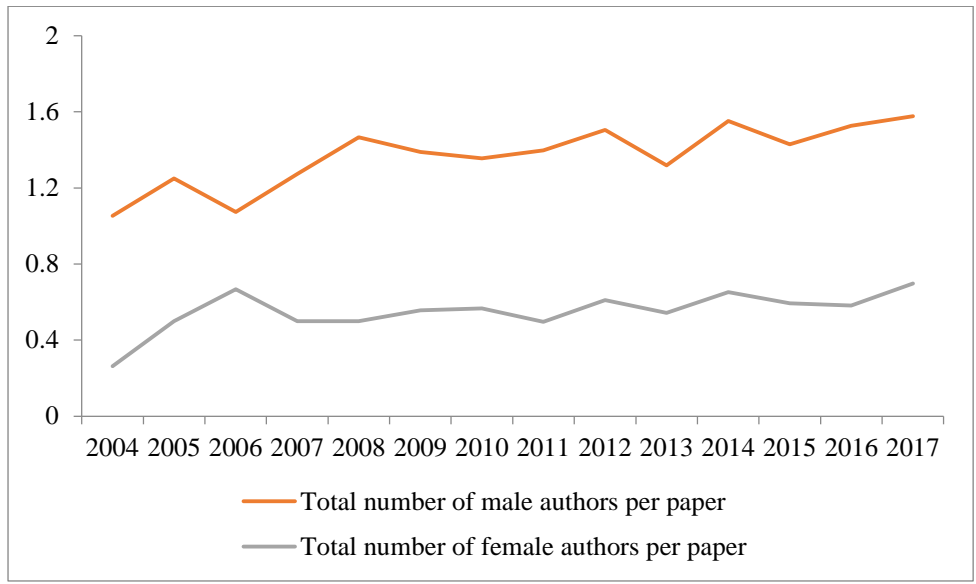

*Source:* Annual Conference on Economic Growth and Development at ISI, Delhi from 2004 to 2017

Figure 3: Composition of authors of accepted papers at ISI-D Conference (single or multi-authored)

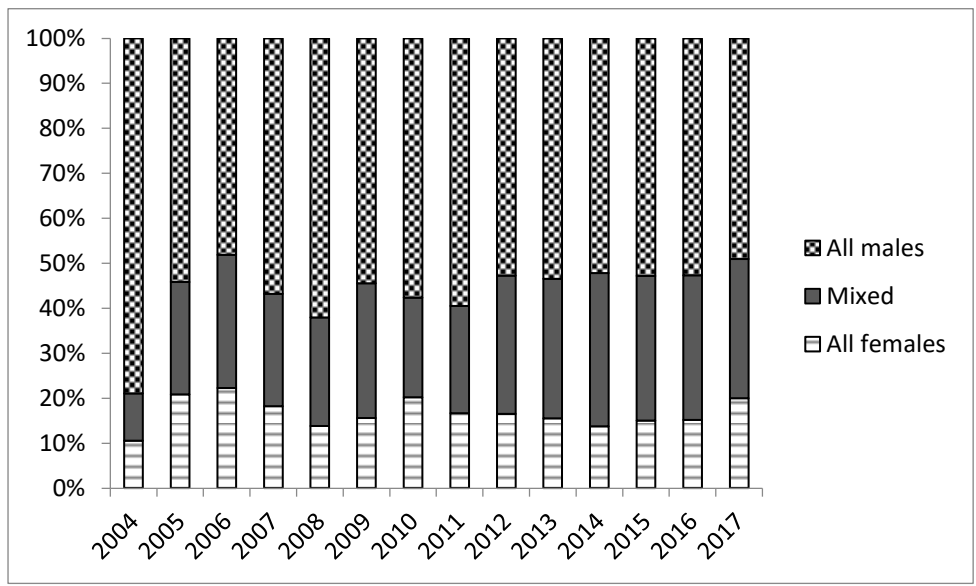

*Source:* Annual Conference on Economic Growth and Development at ISI, Delhi from 2004 to 2017

Figure 4: Composition of authors of published papers (single or multi-authored) in the Indian Journal of Labour Economics

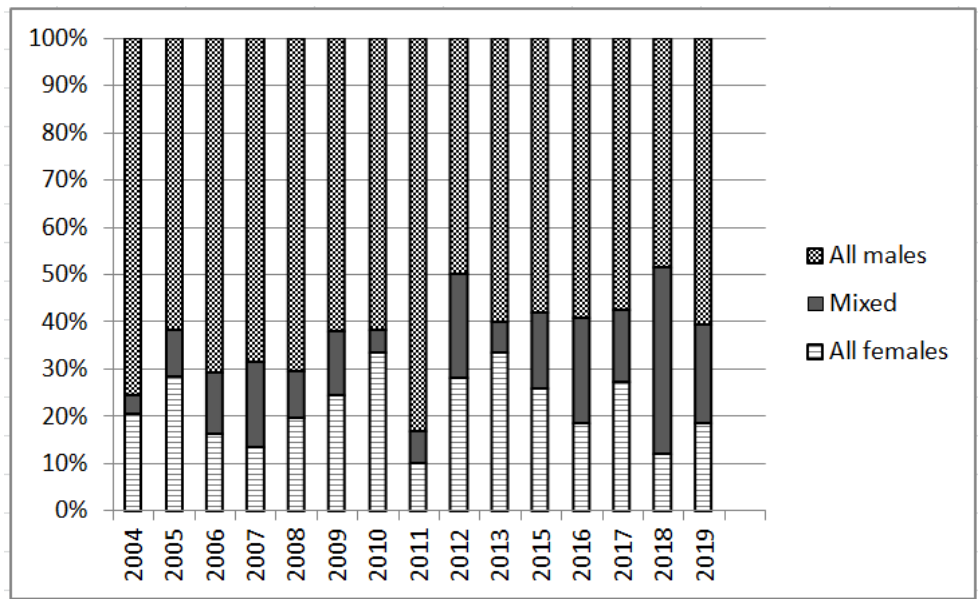

**Tables**

Table 1: State-wise number of institutions included in the sample

| | Combined | | NIRF UNIV 100 | | NIRF 50 MNMT | | NIRF 100 OVERALL | | RePEc 25% | |
|---|---|---|---|---|---|---|---|---|---|---|
| *State* | *n* | *%* | *n* | *%* | *n* | *%* | *n* | *%* | *n* | *%* |
| Andhra Pradesh | 8 | 6.67 | 3 | 5.77 | 2 | 6.25 | 5 | 8.93 | 3 | 7.69 |
| Arunachal Pradesh | 1 | 0.83 | 1 | 1.92 | | | | | | |
| Assam | 4 | 3.33 | 2 | 3.85 | | | 2 | 3.57 | | |
| Bihar | 1 | 0.83 | | | | | 1 | 1.79 | | |
| Chandigarh | 1 | 0.83 | 1 | 1.92 | | | 1 | 1.79 | | |
| Chhattisgarh | 1 | 0.83 | | | 1 | 3.13 | | | | |
| Delhi | 11 | 9.17 | 3 | 5.77 | 4 | 12.5 | 5 | 8.93 | 6 | 15.38 |
| Goa | 2 | 1.67 | 1 | 1.92 | 1 | 3.13 | | | | |
| Gujarat | 5 | 4.17 | | | 2 | 6.25 | | | 4 | 10.26 |
| Haryana | 5 | 4.17 | 2 | 3.85 | 2 | 6.25 | | | 1 | 2.56 |
| Himachal Pradesh | 2 | 1.67 | | | | | 2 | 3.57 | | |
| Jammu & Kashmir | 3 | 2.5 | 2 | 3.85 | | | 1 | 1.79 | 1 | 2.56 |
| Jharkhand | 2 | 1.67 | | | 2 | 6.25 | | | | |
| Karnataka | 9 | 7.5 | 4 | 7.69 | 2 | 6.25 | 3 | 5.36 | 4 | 10.26 |
| Kerala | 6 | 5 | 3 | 5.77 | 1 | 3.13 | 3 | 5.36 | 3 | 7.69 |
| Madhya Pradesh | 2 | 1.67 | | | 1 | 3.13 | 1 | 1.79 | 1 | 2.56 |
| Maharashtra | 8 | 6.67 | 4 | 7.69 | 4 | 12.5 | 3 | 5.36 | 1 | 2.56 |
| Meghalaya | 2 | 1.67 | 1 | 1.92 | 1 | 3.13 | 1 | 1.79 | 1 | 2.56 |
| Mizoram | 1 | 0.83 | 1 | 1.92 | | | | | | |
| Odisha | 4 | 3.33 | 1 | 1.92 | 1 | 3.13 | 3 | 5.36 | | |
| Pondicherry | 1 | 0.83 | | | | | 1 | 1.79 | 1 | 2.56 |
| Punjab | 3 | 2.5 | 3 | 5.77 | | | 2 | 3.57 | | |
| Rajasthan | 3 | 2.5 | 2 | 3.85 | 1 | 3.13 | 2 | 3.57 | 1 | 2.56 |
| Tamil Nadu | 14 | 11.67 | 8 | 15.38 | 2 | 6.25 | 8 | 14.29 | 2 | 5.13 |
| Uttar Pradesh | 8 | 6.67 | 5 | 9.62 | 2 | 6.25 | 5 | 8.93 | 3 | 7.69 |
| Uttarakhand | 2 | 1.67 | | | 1 | 3.13 | 1 | 1.79 | | |
| West Bengal | 11 | 9.17 | 5 | 9.62 | 2 | 6.25 | 6 | 10.71 | 7 | 17.95 |
| **Total** | 120 | 100 | 52 | 100 | 32 | 100 | 56 | 100 | 39 | 100 |

Table 2: Types of institutions included in the sample (as per the UGC classification)

| | NIRF Universities top 100 | NIRF Management Top 50 | NIRF Overall top 100 | RePEc top 25% | All categories |
|---|---|---|---|---|---|
| Central Universities (Public) | 12 | 1 | 9 | 6 | 13 |
| State Universities (Public) | 29 | 0 | 21 | 4 | 34 |
| Deemed to be University | 9 | 3 | 7 | 4 | 12 |
| Institution of National Importance | 0 | 1 | 16 | 4 | 19 |
| IIMs | 0 | 13 | 1 | 7 | 13 |
| Private Universities | 2 | 4 | 2 | 4 | 9 |
| Standalone institutions | 0 | 10 | 0 | 4 | 14 |
| Institutions Recognised by the Universities | 0 | 0 | 0 | 6 | 6 |
| TOTAL | 52 | 32 | 56 | 39 | 120 |

Table 3: Total faculty, female faculty and fraction of female faculty in 'elite' institutions

| Institutions | | Assistant | Associate | Professor | Overall |
|---|---|---|---|---|---|
| NIRF *Overall* top 100 | Female faculty | 59 | 24 | 51 | 136 |
| | Total Faculty | 179 | 77 | 190 | 449 |
| | % | **32.96** | **31.17** | **26.84** | **30.29** |
| NIRF *Universities* top 100 | Female faculty | 51 | 28 | 51 | 132 |
| | Total Faculty | 146 | 72 | 198 | 419 |
| | % | **34.93** | **38.9** | **25.76** | **31.5** |
| NIRF *Management* top 50 | Female faculty | 25 | 15 | 21 | 64 |
| | Total Faculty | 80 | 48 | 79 | 217 |
| | % | **31.25** | **31.25** | **26.58** | **29.49** |
| RePEc top 25% | Female faculty | 64 | 28 | 50 | 142 |
| | Total Faculty | 185 | 85 | 233 | 505 |
| | % | **34.6** | **32.94** | **21.46** | **28.12** |
| All categories combined | Female faculty | 121 | 56 | 100 | 278 |
| | Total Faculty | 357 | 168 | 406 | 939 |
| | % | **33.89** | **33.33** | **24.63** | **29.61** |

Table 4: Fraction of female faculty & Type of the institution (category: NIRF *Overall* Top 100)

| Institutions | Number | *Assistant* | *Associate* | *Professor* | *Overall* |
|---|---|---|---|---|---|
| Central Universities (Public) (n=13) | Female faculty | 16 | 6 | 15 | 39 |
| | Total Faculty | 63 | 25 | 87 | 178 |
| | *%* | *25.4* | *24* | *17.24* | *21.91* |
| State Universities (Public) (n=34) | Female faculty | 31 | 15 | 37 | 83 |
| | Total Faculty | 77 | 40 | 111 | 228 |
| | *%* | *40.26* | *37.5* | *33.33* | *36.4* |
| Deemed to be University (n=12) | Female faculty | 14 | 7 | 12 | 33 |
| | Total Faculty | 35 | 17 | 42 | 94 |
| | *%* | *40* | *41.18* | *28.57* | *35.11* |
| Institution of National Importance (n=19) | Female faculty | 15 | 6 | 6 | 27 |
| | Total Faculty | 53 | 30 | 41 | 124 |
| | *%* | *28.3* | *20* | *14.63* | *21.77* |
| Private Universities (n=9) | Female faculty | 17 | 7 | 6 | 30 |
| | Total Faculty | 35 | 9 | 18 | 62 |
| | *%* | *48.57* | *77.78* | *33.33* | *48.39* |
| Indian Institute(s) of Management (IIMs) (n=13) | Female faculty | 12 | 7 | 6 | 25 |
| | Total Faculty | 39 | 24 | 37 | 104 |
| | *%* | *30.77* | *29.17* | *16.22* | *24.04* |
| Other Standalone institutions (n=14) | Female faculty | 11 | 5 | 8 | 25 |
| | Total Faculty | 30 | 15 | 29 | 76 |
| | *%* | *36.67* | *33.33* | *27.59* | *32.89* |
| Institutions recognised by university (n=6) | Female faculty | 5 | 3 | 10 | 16 |
| | Total Faculty | 25 | 8 | 41 | 73 |
| | *%* | *20* | *37.5* | *24.39* | *21.92* |

Table 5: Fraction of women in Economics master's programmes (overall)

|  | Enrolment in PG (%) | Completed PG (%) |
|---|---|---|
| 2011-12 | 52 | 52 |
| 2012-13 | 54 | 56 |
| 2013-14 | 54 | 56 |
| 2014-15 | 55 | 53 |
| 2015-16 | 56 | 57 |
| 2016-17 | 57 | 59 |
| 2017-18 | 56 | 60 |

*Source: All India Survey of Higher Education (MHRD, GoI) of the respective years*

Table 6: Fraction of women in Economics master's programmes (CDS)

| Year | Total | Female | % Female |
|---|---|---|---|
| 2012-14 | 15 | 10 | 66.7 |
| 2013-15 | 17 | 5 | 29.4 |
| 2014-16 | 17 | 10 | 58.8 |
| 2015-17 | 20 | 10 | 50 |
| 2016-18 | 21 | 13 | 61.9 |
| 2017-19 | 21 | 8 | 38.1 |
| 2018-20 | 18 | 7 | 38.9 |
| Overall | 129 | 63 | 48.8 |

*Source: Information available on the website*

Table 7: Fraction of women in Economics Master's programs (IGIDR)

| Year | Total | Female | % Female |
|---|---|---|---|
| 2010 | 28 | 17 | 60.7 |
| 2011 | 21 | 6 | 28.6 |
| 2012 | 25 | 15 | 60 |
| 2013 | 20 | 10 | 50 |
| 2014 | 24 | 17 | 70.8 |
| 2015 | 22 | 14 | 63.6 |
| 2016 | 19 | 12 | 63.2 |
| 2017 | 28 | 13 | 46.4 |
| 2018 | 29 | 16 | 55.2 |
| Overall | 216 | 120 | 55.6 |

*Source: Response from request under RTI Act*

Table 8: Fraction of women in Economics master's programmes (ISI – Delhi and Kolkata combined)[xx]

| Year | Total | Females | % Female |
|---|---|---|---|
| 1998 | 15 | 5 | 33.3 |
| 1999 | 14 | 3 | 21.4 |
| 2000 | 20 | 7 | 35 |
| 2001 | 13 | 7 | 53.8 |
| 2002 | 22 | 9 | 40.9 |
| 2003 | 22 | 9 | 40.9 |
| 2004 | 19 | 8 | 42.1 |
| 2005 | 20 | 11 | 55 |
| 2006 | 17 | 6 | 35.3 |
| 2007 | 21 | 5 | 23.8 |
| 2009 | 31 | 14 | 45.2 |
| 2010 | 39 | 17 | 43.6 |
| 2011 | 27 | 12 | 44.4 |
| 2012 | 33 | 15 | 45.5 |
| 2013 | 23 | 11 | 47.8 |
| 2014 | 17 | 6 | 35.3 |
| 2015 | 21 | 13 | 61.9 |
| 2016 | 23 | 10 | 43.5 |
| 2017 | 22 | 9 | 40.9 |
| 2018 | 27 | 8 | 29.6 |
| Overall | 446 | 185 | 41.5 |

*Source: Information available on the website and placement brochures*

Table 9: Fraction of women in Economics master's programmes (**Second Year**, Delhi School of Economics)

| Year | Total | Female | % Female |
|------|-------|--------|----------|
| 2011 | 99    | 48     | 48.5     |
| 2017 | 144   | 89     | 61.8     |

*Source: Annual Reports, University of Delhi*

Table 10: Fraction of women in MPhil and PhD in Economics (overall)

|  | *Enrolled (%)* | | *Earned (%)* | |
| --- | --- | --- | --- | --- |
|  | **PhD** | **MPhil** | **PhD** | **MPhil** |
| 2011-12 | 35 | 48 | 32 | 51 |
| 2012-13 | 42 | 49 | 42 | 53 |
| 2013-14 | 41 | 52 | 37 | 54 |
| 2014-15 | 41 | 57 | 40 | 58 |
| 2015-16 | 45 | 56 | 38 | 59 |
| 2016-17 | 46 | 58 | 43 | 60 |
| 2017-18 | 47 | 63 | 40 | 62 |

*Source: All India Survey of Higher Education, MHRD*

Table 11: Fraction of women earning doctorate in Economics in the US

| Year | Male | Female | Total | % Female |
|------|------|--------|-------|----------|
| 1997 | 31 | 20 | 51 | 39.2 |
| 1998 | 39 | 19 | 58 | 32.8 |
| 1999 | 31 | 15 | 46 | 32.6 |
| 2000 | 21 | 18 | 39 | 46.1 |
| 2001 | 25 | 18 | 43 | 41.9 |
| 2002 | 25 | 20 | 45 | 44.4 |
| 2003 | 19 | 14 | 33 | 42.4 |
| 2004 | 22 | 16 | 38 | 42.1 |
| 2005 | 31 | 17 | 48 | 35.4 |
| 2006 | 30 | 21 | 51 | 41.2 |
| 2007 | 26 | 17 | 43 | 39.5 |
| 2008 | 22 | 21 | 43 | 48.8 |
| 2009 | 36 | 24 | 60 | 40 |
| 2010 | 28 | 21 | 49 | 42.9 |
| 2011 | 33 | 33 | 66 | 50 |
| 2012 | 26 | 36 | 62 | 58.1 |
| 2013 | 23 | 18 | 41 | 43.9 |
| 2014 | 25 | 16 | 41 | 39 |
| 2015 | 16 | 19 | 35 | 54.3 |
| 2016 | 20 | 23 | 43 | 53.5 |
| 2017 | 22 | 23 | 45 | 51.1 |

*Source: National Science Foundation, National Centre for Science and Engineering Statistics, Survey of Earned Doctorates*

Table 12: Fraction of women among faculty members across locations

|  |  | Metro Locations | Non-metro locations |
|---|---|---|---|
| Assistant Professor | Total faculty | 57 | 64 |
|  | Female Faculty | 156 | 201 |
|  | *%* | ***36.5*** | ***31.8*** |
| Associate Professor | Total faculty | 28 | 28 |
|  | Female Faculty | 73 | 95 |
|  | *%* | ***38.4*** | ***29.5*** |
| Professor | Total faculty | 63 | 37 |
|  | Female Faculty | 220 | 186 |
|  | *%* | ***28.6*** | ***19.9*** |
| Overall | Total faculty | 149 | 129 |
|  | Female Faculty | 453 | 486 |
|  | *%* | ***32. 9*** | ***26.5*** |

**Endnotes**

[i] Auriol et al. (2019) also use RePEc in similar work on European institutions.

[ii] The list of institutions is available here: https://ideas.repec.org/top/top.india.html

[iii] Details are available on RePEc website.

[iv] The list of institution in each of the four categories is available on request.

[v] Examples in case of management institutions include departments or centres of public policy (e.g. Indian Institute of Management (IIM) Bangalore or Indian Institute of Technology (IIT) Delhi) or Centre for Management of Agriculture in IIM Ahmedabad.

[vi] Details are provided in Table A2 of the online annexure.

[vii] Available at: https://www.isid.ac.in/~epu/acegd2019/past-conferences.html (accessed on September 15, 2019)

[viii] As a result, we can't assess share of women at the submission stage, the acceptance stage or at any stage from acceptance to the freezing of final schedule. This raises the possibility of our results on proportion of women at ISI conference being understated if one assumes that women are more likely to drop out before final schedule is decided.

[ix] Reports and data available at: http://aishe.nic.in/aishe/home

[x] https://www.isical.ac.in/~deanweb/ALU_YEAR.HTM

[xi] Sabharwal et al. (2019) also find this in their work which is based on administrative data and interviews with faculty in institutions carrying out research in STEM.

[xii] Results are graphically presented in figure F1 in the online annexure.

[xiii] These institutions are also part of our sample of 'elite' institutions.

[xiv] In fact, there is also a substantial gap between per cent of women enrolled in the PhD programme and per cent of women who eventually earn the PhD.

[xv] These cities are Bengaluru, Chennai, Hyderabad, Kolkata, Mumbai, and New Delhi (including and excluding National Capital Region). We show results separately for Delhi and NCR (including Delhi).

[xvi] Women with children publish significantly fewer papers than men with children while no gap exists between men and women without children (Ginther et al. 2017).

[xvii] Note that discussion here is aimed to be generic and not specific to the conference mentioned in the paper. In fact, to our knowledge, the ISI conference organizers are quite accommodative of genuine constraints of all the participants.

[xviii] With the Covid-19 pandemic, virtual conferencing might be the only option at least for some time.

[xix] https://www.aeaweb.org/resources/best-practices

[xx] We do not have information on total number of students and their gender for ISI-Kolkata 2014 onward.